\documentclass[11pt,twoside]{article}
\usepackage{newpasp}

\index{Revaz, Y.}
\index{Pfenniger, D.}

\begin{document}

\title{N-body simulations of warped galaxies}
\author{Y. Revaz \& D. Pfenniger}
\affil{Geneva Observatory, CH-1290 Sauverny, Switzerland}


\begin{abstract}

Two methods generating warped galaxies with N-body simulations are presented. One uses an
external potential as a disturber while the other is based on material accretion. The 
results of both methods are compared. A particular attention is given on the shape of the
line of node (LON). 

\end{abstract}




\section{Introduction}

The understanding of warped galaxies has been enriched those last years by
ideas such as galactic infall (Ostriker \& Binney \ 1989; Jiang \& Binney\ 1999) 
or dynamical friction between the halo and the disk (Debattista \& Sellwood 1999). 
In spite of the fact that the evidences for non-self-gravitating HI disks
are actually weak (Arnaboldi et al. 1997; Becquaert \& Combes 1997), 
most of works on the subject are built on the assumption of a disk embedded in a fat spheroidal or 
triaxial halo.

Through N-body simulations, this work presents two methods of getting 
warped self-gravitating galactic disks. 
The first one imports inclined angular momentum through material accretion and 
the second one uses an outer warped potential. 

\section{Models and simulations}

The N-body disk is based on a superposition
of three Miyamoto-Nagai potentials~:
        \begin{equation}
	\label{miyam}
        \Phi(R,\phi,z) = -\sum_{i=1}^{3} \frac{GM_i}
        {\sqrt{R^2+\left(a_i+\sqrt{z^2 + b_i^2}\right)^2}}
	\end{equation}
with the components representing a bulge, a visible disk and a gas disk.

\subsection{External potential}

In order to warp the galaxy, an initially flat disk is embedded in a non-rotating 
warped potential.
This outer potential whith a mass equivalent to the disk is obtained by replacing 
$z$ by $z-w R^2 \cos\phi$ in the disk potential of equation~\ref{miyam}. 
$w$ is a parameter determining the amplitude of the warp.  To drive
adiabatically the disk, the outer potential is flat at the beginning of the simulation 
and warps slowly during the perturbation time. Afterwards, the potential is
removed and the disk evolves freely. 

After a few tens of Myr, the inner part of the disk tilts while the outer remains flat. 
At $t\cong 750\,\textrm{Myr}$ the whole disk has warped in order to follow the outer
potential and the LON is observable as a leading spiral. 
The maximum amplitude of the warp at $R=30\,\textrm{kpc}$ is about $20\,\textrm{kpc}$.
After the perturbation,  the deformation of the isolated disk 
decreases from the center to the outside and the LON becomes
straight. This last step needs about $1\,\textrm{Gyr}$.

\subsection{Accretion}

In this simulation, the warp is generated by material accretion. This idea has been
suggested by Ostriker \& Binney \ (1989) and tested by Jiang \& Binney\ (1999). 
The originality of our simulation is the absence of halo.
The infall of material
is obtained by injecting particles in an inclined tork. Its velocity corresponds to the 
velocity of the  rotation curve at an equivalent radius.
The total injected mass
represent 100\,\% of the mass of the initial galactic disk. 

The disk needs about $1\,\textrm{Gyr}$ in order to react to the accretion and the 
deformation will need one more Gyr to propagate from the outside to the center.
The maximum
amplitude of the warp at $R=30\,\textrm{kpc}$ is about $10\,\textrm{kpc}$. A straight
retrograde rotating LON is associated to the warp. Once the accretion is ended the disk
needs about $500\,\textrm{Myr}$ to become flat again.

\section{Discussion}

The perturbation type determines strongly the evolution and shape of the warped disk. 
In the first simulation,
the deformation occurs  from the inner regions to the outer whereas it occurs from the
outer to the inner in the second one. This has two observed effects : 

The shape of the LON is significantly different. 
A \textbf{leading LON} is generated in presence of an outer potential, 
which can be assimilated to a warped heavy DM disk. 
A similar result has been found by Debattista \& Sellwood (1999) in the case of 
dynamical friction between a spherical misaligned rotating halo and a disk.
In contrary, a \textbf{straight rotating LON} is generated with accretion.

The birth of the warp is also quite different. In case of our outer potential, the inner
regions feel its gravity faster and tip before the outer ones. A large abrupt warp can be
quickly observed, even if the outer regions lie always on the initial plane.  This effect
is not present in case of accretion, where the inner regions remain flat while the outer
ones tip slowly. 


\end{document}